\documentclass[12pt]{article}
\usepackage{epsfig}

\def\beq{\begin{equation}}
\def\eeq#1{\label{#1}\end{equation}}
\def\eeqn{\end{equation}}

\def\beqa{\begin{eqnarray}}
\def\eeqa#1{\label{#1}\end{eqnarray}}
\def\eeqan{\end{eqnarray}}

\let\bar=\overbar

\def\O{{\cal O}}

\def\Dslash{\not{\hbox{\kern-4pt $D$}}}
\def\dslash{\not{\hbox{\kern-2pt $\del$}}}

\def\BR{\mbox{\rm BR}}

\def\msb{{\bar{\ssstyle M \kern -1pt S}}}

\def\Title#1{\begin{center} {\Large {\bf #1} } \end{center}}

\begin{document}

\Title{Heavy B Hadrons}

\bigskip\bigskip

\begin{raggedright}  

{\it Stefano Giagu \\
for the CDF and D\O\ Collaborations\\
Sapienza Universit\`a di Roma and INFN Roma, 00185-Roma, IT}
\bigskip\bigskip
\end{raggedright}

\section{Introduction}

The CDF and D\O\ experiments have successfully collected data since start of the Run II at the Tevatron Collider in 2001. 
The large B-meson production cross-section and the possibility to produce all kind of B hadron states, opened to the two collaborations
the possibility to study with high precision the tiny effects of \textsf{CP}-violation in the Heavy B hadrons system, and to search for new physics effects in 
rare decays, in a way unavailable to the previous generation experiments. 
A new and largely unknown sector of the Heavy Flavor physics, complementary to the one already tested with precision at the B-factories, as recently 
pointed out by I. Bigi~\cite{bigi}, has begun to be explored in search of possible signs of new physics. In this short note a selection of the most recent results 
on heavy B hadrons (mostly $B_s$ mesons) from the Fermilab Tevatron, and from the Belle experiment running at the $\Upsilon(5S)$ are reviewed.

\section{\textsf{CP} Violation in $B_s$ mesons at Tevatron}

In the neutral $B_s$ system, the \textsf{CP} asymmetry in $B^0_s \to J/\psi \phi$ decay play the analogous role of the $B^0 \to J/\psi K_s^0$ for the 
$B_d$ system.
Decays of the $B_s$ meson via $b\to c\bar{c}s$ transitions in fact can be used to probe, via interference effects in the mixing and decay amplitudes 
of the process, the $\beta_s$ angle of the ÒsquashedÓ (bs) unitarity triangle, defined as $\beta_s = arg \left ( -{V_{ts}V^*_{tb}}/{V_{cs}V^*_{cb}} \right )$.
An important difference with respect to the $B_d$ system is that in the Standard Model $\beta_s$ is expected to be very small ($\sim 0.02$), 
making the measurement of \textsf{CP} asymmetry in  $B_s^0 \to J/\psi \phi$ decay very sensitive to new possible physics effects in the mixing phase 
of the $B_s^0-\bar{B}_s^0$ system.

CDF and D\O\ experiments have been able, for the first time, to perform a search for \textsf{CP} violation in the neutral $B_s$ meson system, by measuring the time-dependent \textsf{CP} asymmetry in the $B^0_s \to J/\psi \phi$ decay mode. 
The vector-vector final state $J/\psi\phi$ contains mixtures of polarization amplitudes: the \textsf{CP}-odd $A_{\perp}$, and the \textsf{CP}-even $A_0$ and $A_{||}$ amplitudes. These terms need to be disentangled, using the angular distributions, in order to extract $\beta_s$, and their interference provides additional sensitivity~\cite{dighe}.

\begin{figure}[htb]
\centering{
\epsfig{figure=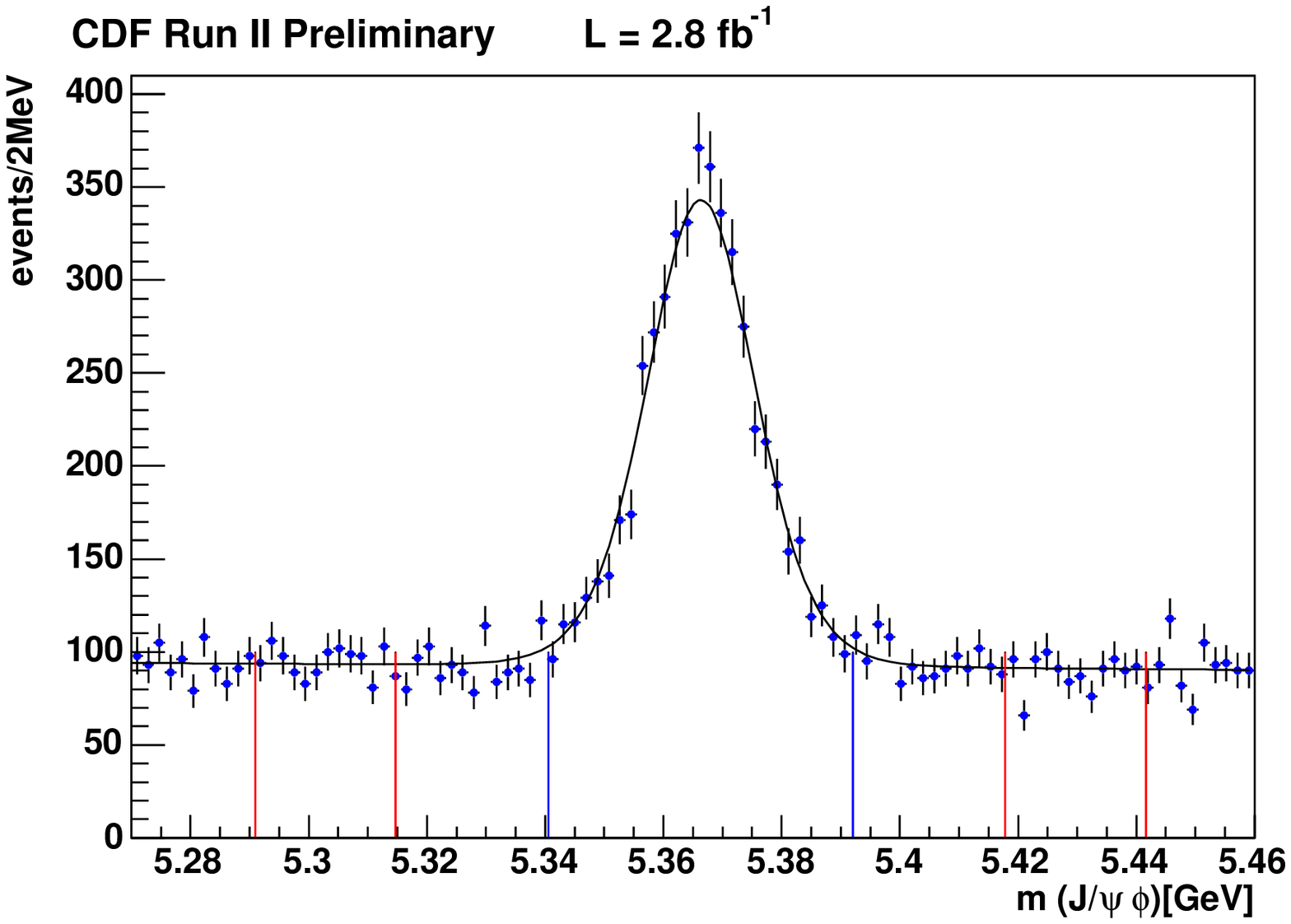,width=0.46\textwidth,height=0.27\textheight} \hspace{0.3cm}
\epsfig{figure=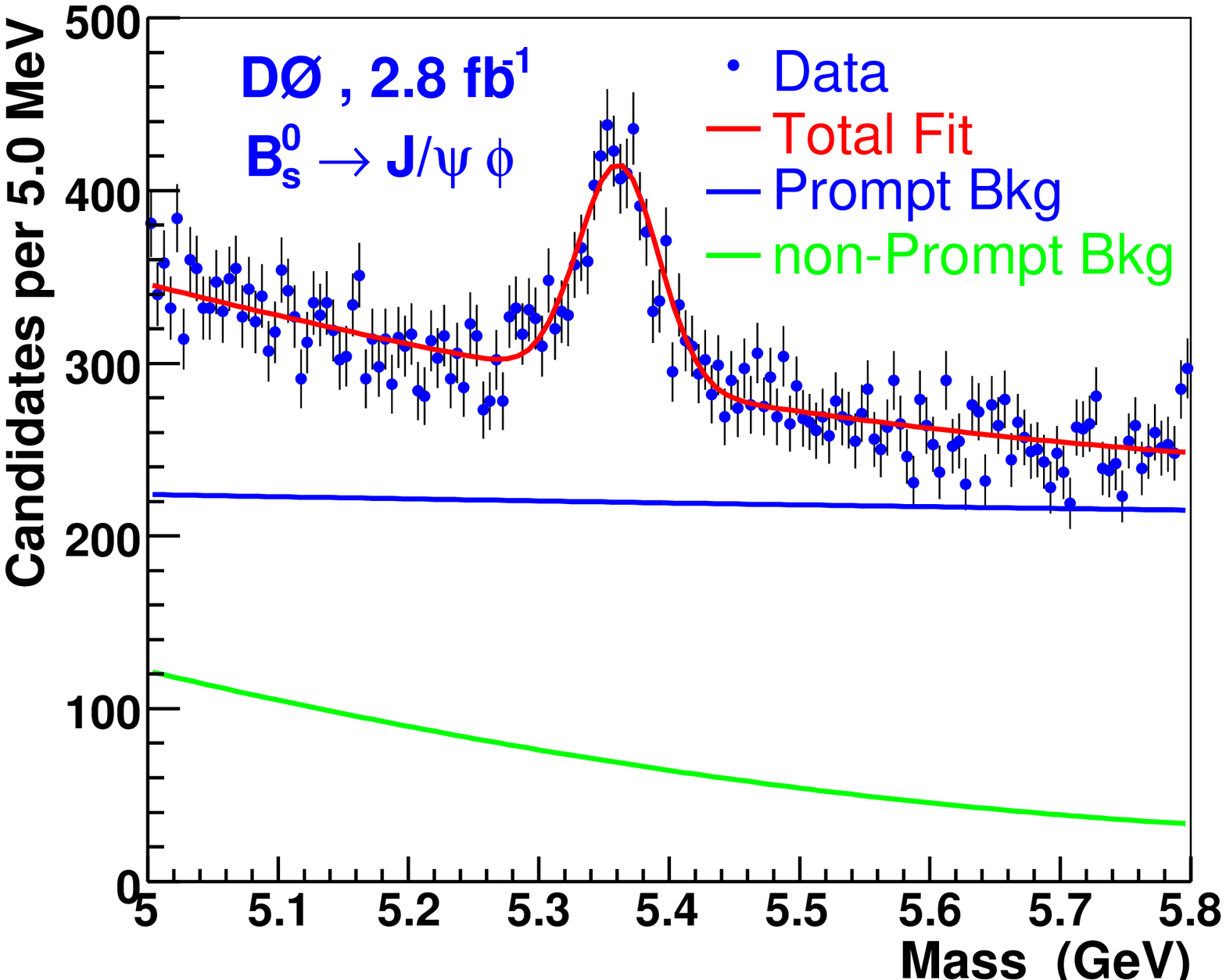,width=0.43\textwidth,height=0.25\textheight}
\caption{Data samples for the CDF and D\O\ analyses of $B^0_s\to J\psi\phi$ decay.  The CDF experiment reports $3166\pm56$ events, while
the D\O\ experiment reports $1967\pm65$ events.}
\label{fig:beta1}
}
\end{figure}
Very recently the CDF collaboration presented a new updated flavor-tagged, time-dependent, analysis, based on 2.8 fb$^{-1}$ of integrated luminosity~\cite{cdfbeta0}, that  supersedes results from the previous 1.7 $fb^{-1}$ untagged analysis~\cite{cdfbeta1}, and 1.35 fb$^{-1}$ flavor-tagged, time-dependent analysis~\cite{cdfbeta2}.
The D\O\ collaboration result is also based on a 2.8 $fb^{-1}$ sample of flavor-tagged data, and is published in~\cite{d0beta}.
The $B^0_s \to J/\psi \phi$ signals from the two experiments are shown in Fig.~\ref{fig:beta1}. 
\begin{figure}[htb]
\centering{
\epsfig{figure=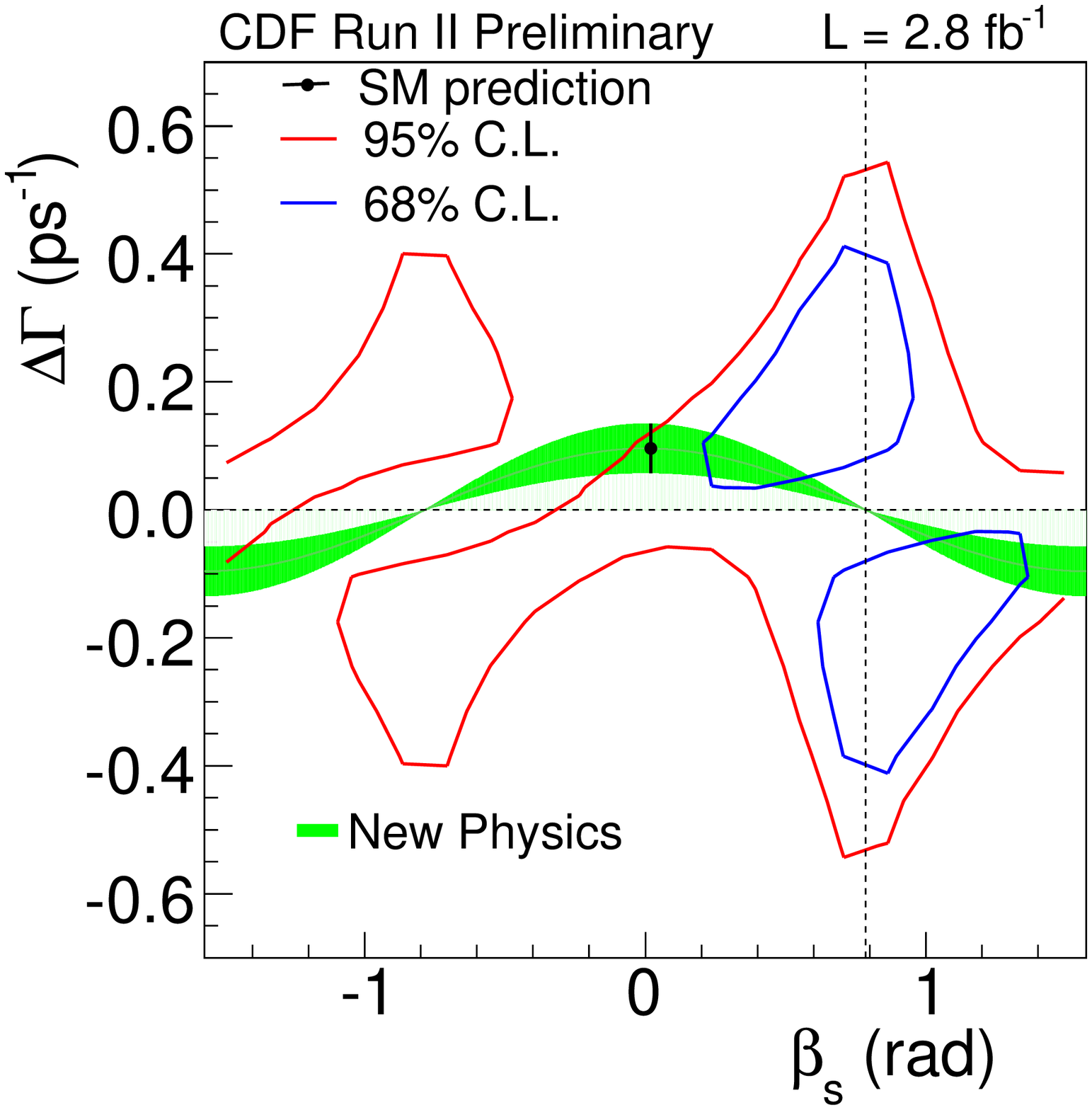,width=0.495\textwidth,height=0.28\textheight} \hspace{0.4cm}
\epsfig{figure=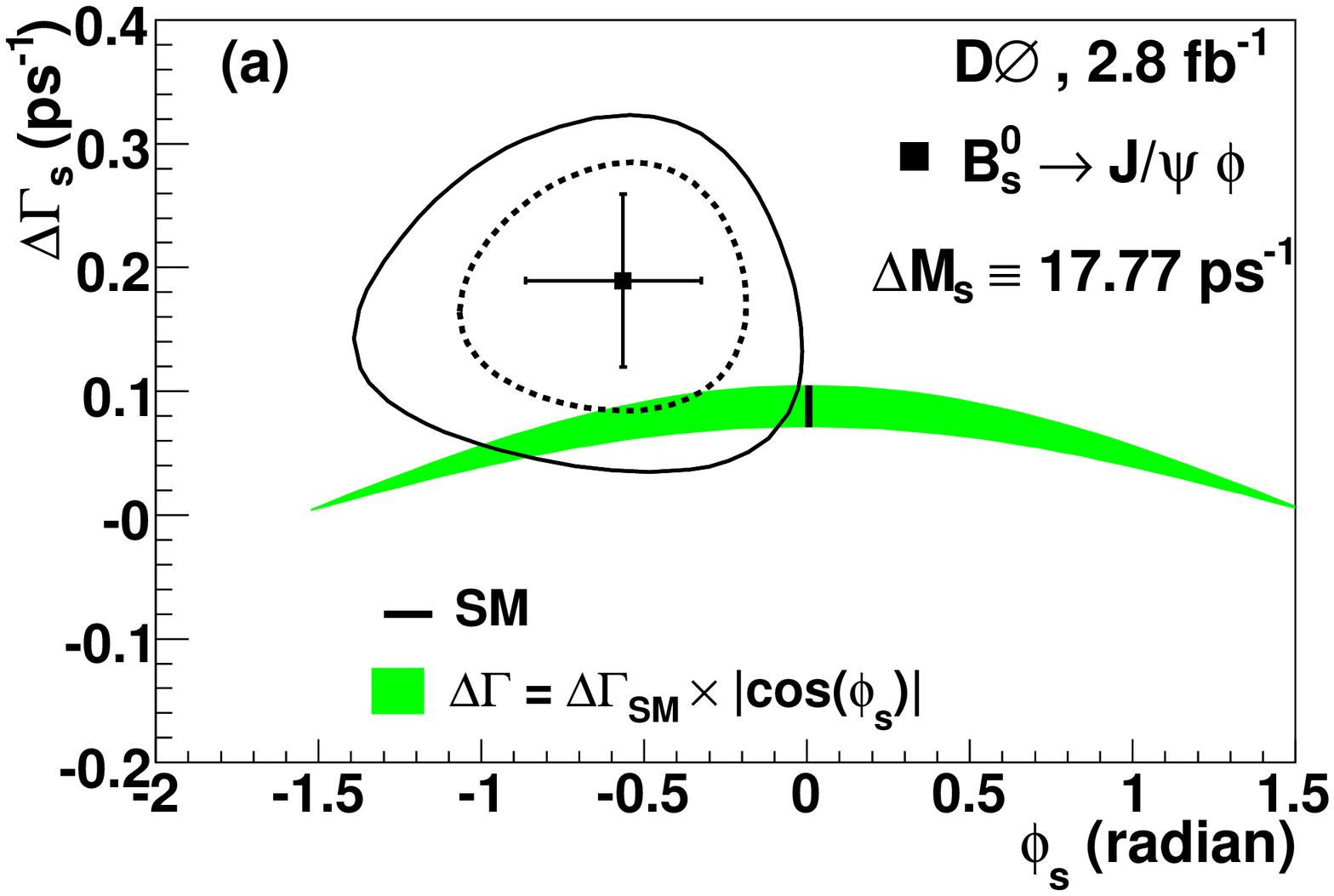,width=0.445\textwidth,height=0.27\textheight}
\caption{Confidence regions in the space of parameters $\Delta\Gamma_s$ and $\beta_s$ for the CDF (left), and D\O\ (right) analyses.
The green band corresponds to new physics models, as described in the text.}
\label{fig:beta2}
}
\end{figure}

In the CDF analysis two different fits have been performed, one without using flavor tagging and assuming the Standard Model, which fixes $\beta_s = 0$, 
and a second one with flavor-tagging and letting $\beta_s$ to be free. Assuming no \textsf{CP} violation ($\beta_s = 0$) the fit allows to simultaneously measure the decay width difference $\Delta\Gamma_s$ and 
the average lifetime of the $B_s$ meson. The results are reported in Table~\ref{tab:beta1}. 
\begin{table}[h]
\centering{
\caption{\label{tab:beta1} Standard Model fits in the CDF tagged analysis.}
\vspace{0.3cm}
\begin{tabular}{|l|c|}
\hline
Parameter & CDF measurement (untagged) \\\hline
$c\tau_s = 2c/(\Gamma_H + \Gamma_L)$ & $459\pm12 (stat)\pm 3 (syst)~\mu$m \\\hline
$\Delta\Gamma_s = \Gamma_L - \Gamma_H$ & $0.02\pm0.05 (stat)\pm0.01 (syst)~$ps$^{-1}$ \\\hline
$|A_0|^2$ & $0.508\pm0.024\pm0.008$ \\\hline
$|A_{||}(0)|^2$ & $0.241\pm0.019\pm0.007$ \\\hline
\end{tabular}
}
\end{table}

It is worth noticing here that the average $B_s$ lifetime measurement is consistent with the HQET expectation of equal lifetimes for the $B_s$ and $B_d$ mesons.
Table~\ref{tab:beta1} reports also the measured polarization amplitudes for the  $B_s^0 \to J/\psi \phi$ decay, that are consistent with those
measured in the similar $B_d$ decay $B^0 \to J/\psi K^{*0}$~\cite{b0ampli}.
For the \textsf{CP} fit, CDF does not report point estimates for any of the physics parameters, providing instead the confidence region in the 
($\beta_s, \Delta\Gamma_s$) plane shown in Fig.~\ref{fig:beta2} (left), computed from Monte Carlo using Feldman-Cousins method for confidence intervals. Also shown in the same figure, is the theoretical expectation from the Standard Model (black point), and in presence of new physics (green band).
The Standard Model {\it p}-value calculated using the likelihood ratio is of 7\%, corresponding to 1.8 Gaussian standard deviations. 

Treating $\Delta\Gamma_s$ as a nuisance parameter, CDF reports confidence intervals for $\beta_s$, and find that $\beta_s$ is within 
[0.28, 1.29] at  68\% C.L., and within [-$\pi$/2, -1.45] $\cup$ [-1.01, -0.57] $\cup$ [-0.13, $\pi$/2] at 95\% CL.

To remove the two-fold ambiguity in the likelihood of the time-dependent, flavor tagged analysis, D\O\ constrained the strong phases 
 of the helicity amplitudes in the $B_s^0 \to J/\psi \phi$ decay to the world average values for the $B^0 \to J/\psi K^{*0}$ decay, measured at the 
 B-factories~\cite{hfag1}. Some justification of the constraint used has been recently showed in~\cite{theobetas}.
 
Results of the constrained fits are shown Table~\ref{tab:beta2}. 
\begin{table}[h]
\centering{
\caption{\label{tab:beta2} CDF results from tagged $B_s^0 \to J/\psi\phi$ analysis.}
\vspace{0.3cm}
\begin{tabular}{|l|c|c|c|}
\hline
Parameter &\textsf{CP} Fit & SM Fit  & NP Fit \\
   & ($\phi_s$ floating) & ($\phi_s=0$) & ($\Delta\Gamma_s = 2|M_{12}^s|\cos\phi_s$ constraint) \\\hline
$\tau_s$ (ps)                                  & $1.52\pm0.06$ & $1.53\pm0.06$ & $1.49\pm0.05$ \\\hline
$\Delta\Gamma_s$ (ps$^{-1}$) &  $0.19\pm0.07$ & $0.14\pm0.07$ & $0.083\pm0.018$ \\\hline 
$A_{\perp}(0)$                              &  $0.41\pm0.04$ & $0.44\pm0.04$ & $0.45\pm0.03$ \\\hline
$|A_0(0)|^2-|A_{||}(0)|^2$                & $0.34\pm0.05$ & $0.35\pm0.04$ & $0.33\pm0.04$ \\\hline
$\phi_s = -2\beta_s$                    & $-0.57^{+0.24}_{-0.30}$ & fixed ($-0.04$) & $-0.46\pm0.28$ \\\hline
\end{tabular}
}
\end{table}
In this case three types of fit have been performed: a Standard Model fit which fixes $\beta_s$ to its expected value, a  \textsf{CP} fit with $\beta_s$ floating, and in addition, a 
\textsf{CP} fit with the further constraint that $\Delta\Gamma_s = 2|\Gamma_{12}^s|\cos\phi_s$, where $\phi_s = arg (-M_{12}^s/\Gamma_{12}^s)$, is the 
mixing phase of the $B_s$ system\footnote{In the discussion the approximation $\phi_s \sim -2\beta_s$ has been made. This is a reasonable approximation since, although the equality does not hold in the Standard Model, both are much smaller than the current experimental resolution, whereas new physics contributions add a phase $\phi_{NP}$ to $\phi_s$  and subtract the same phase from $2\beta_s$, so that the approximation remains valid.}, and $M_{12}^s$ is the off-diagonal element 
of the mass matrix governing the flavor oscillations in the $B_s$ system (related to the mixing frequency by $\Delta m_s = 2|M_{12}^s|$).
The average lifetime $\tau_s$, and the decay amplitudes are consistent with expectations and with the CDF measurements. Confidence 
regions in the ($\phi_s( = -2\beta_s), \Delta\Gamma_s$) space are shown in Fig.~\ref{fig:beta2} (right). The point estimate for the 
CP violation phase and $\Delta\Gamma_s$, obtained by D\O\ are: $\phi_s = -0.57^{+0.24}_{-0.30} (stat) ^{+0.07}_{-0.02} (syst)$ and 
$\Delta\Gamma_s = 0.19\pm0.07 (stat)^{+0.02}_{0.01} (syst)~$ ps$^{-1}$. The Standard Model {\it p}-value is of 6.6\%, and the fluctuation respect to 
the Standard Model goes in the same direction as CDF.

Experimental sensitivity to new physics effects on the $B_s$ mixing phase can also be obtained from charge asymmetry measurements in semi-leptonic 
$B_s$ decays. The semi-leptonic asymmetry is in fact related to the mixing phase by the relation: 

\begin{eqnarray*} 
A^{s}_{SL}=\frac{N(\bar{B}_{s}\to f) - N(B_{s}\to \bar{f})}{N(\bar{B}_{s}\to f) + N(B_{s}\to \bar{f})} \sim \frac{\Delta\Gamma_{s}}{\Delta m_{s}}\tan \phi_{s},
\label{eq:asl_def}
\end{eqnarray*}

where $f$ corresponds to direct $B_{s}$ decays $B_{s}\to f$ (e.g. $D^{-}_{s}l^{+}\nu_{l}$). 

The Standard Model prediction for the semileptonic asymmetry in $B_{s}$ decays is very small, at the level of few units in 10$^{-5}$~\cite{lenz}.

At the Tevatron the semileptonic asymmetry has been measured both in inclusive di-muon samples, where
$ A_{SL}^{s}\sim \frac{ N_{\mu^{+}\mu^{+}} - N_{\mu^{-}\mu^{-}} }{ N_{\mu^{+}\mu^{+}}+ N_{\mu^{-}\mu^{-}} } $,
or by using the sequential decays sample $B_s^0 \to \mu\nu D_{s}$. The first method has very high statistical accuracy, but requires knowledge 
of asymmetries of other contributing processes in addition to the detector charge asymmetries. The 
second method has less statistical power but ensures that the major contribution to the asymmetry 
comes from the $B_{s}$ decays. Combining the two measurements D\O\  obtains: $A_{SL}^{s}=0.0001\pm 0.0090$~\cite{d0asl}, 
while the CDF result based on 1.6 fb$^{-1}$ di-muon pairs, is:  $A_{SL}^{s}=0.020\pm 0.028$~\cite{cdfasl}.
At the current level of precision $A^s_{SL}$ is not able to provide powerful constraints on new physics contributions on 
the mixing phase.

\begin{figure}[htb]
\centering{
\epsfig{figure=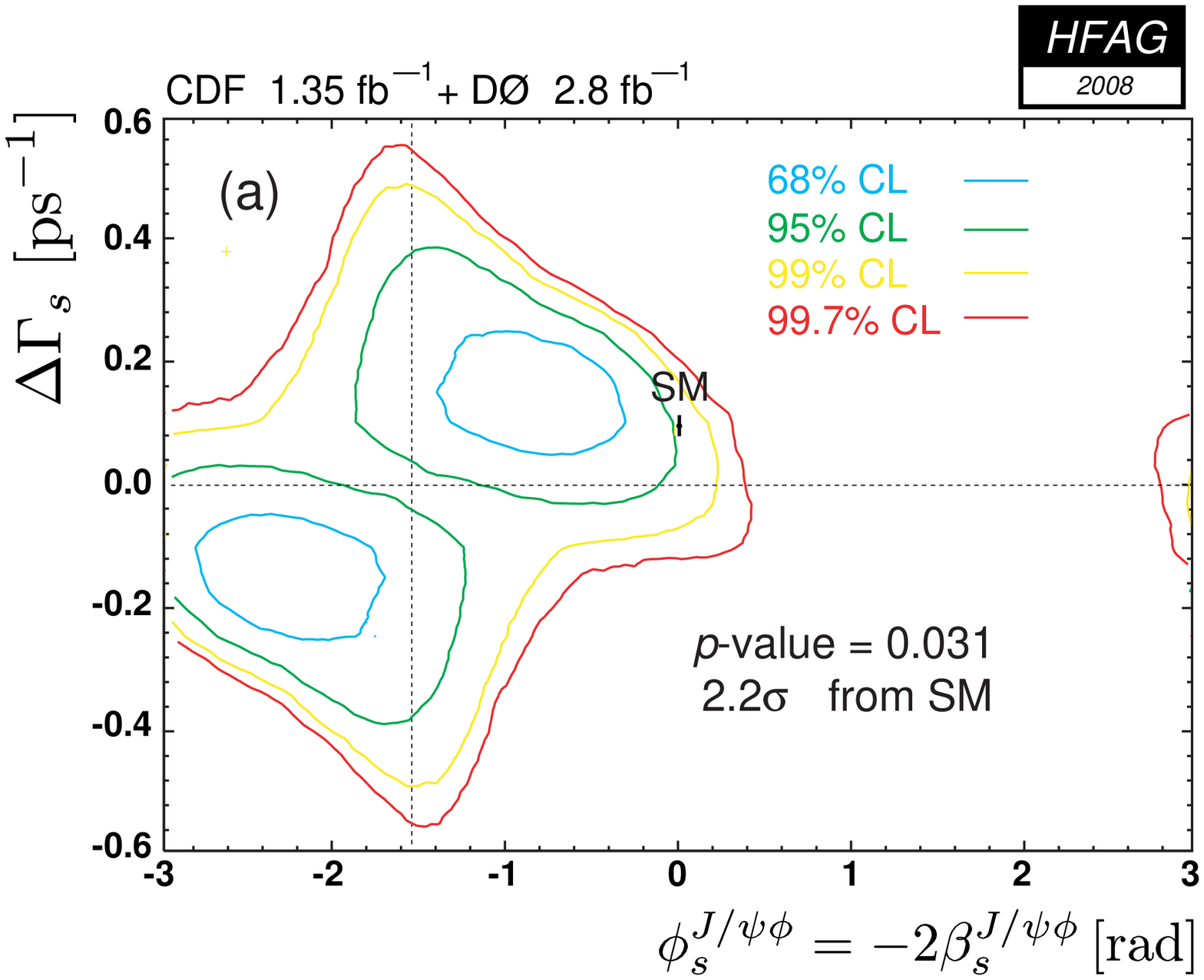,width=0.47\textwidth,height=0.28\textheight} \hspace{0.3cm}
\epsfig{figure=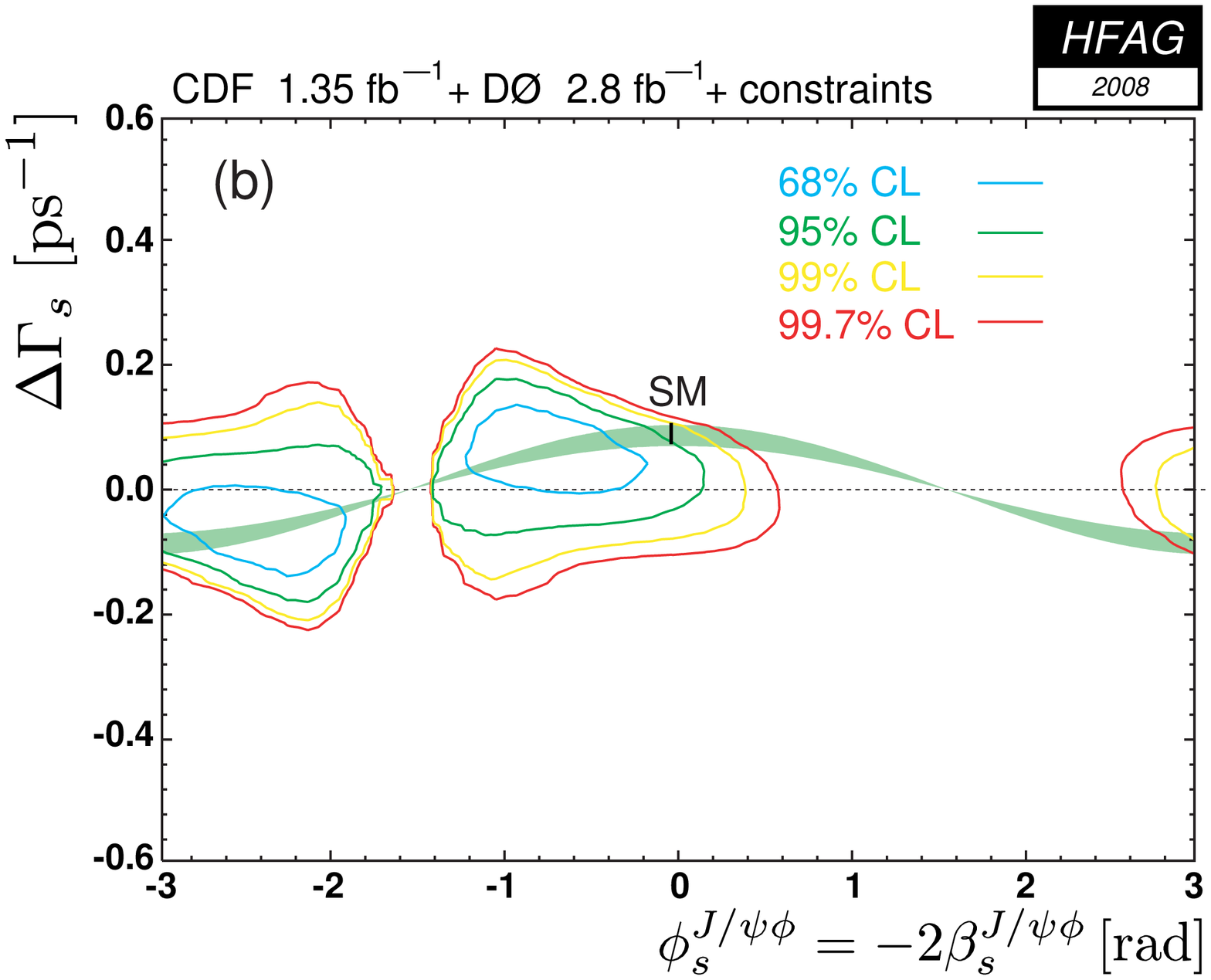,width=0.47\textwidth,height=0.28\textheight}
\caption{HFAG Combinations of the CDF and D\O\  $B_s^0\to J/\psi\phi$ tagged analyses. The plot on the right use additional experimental input 
 from the CDF and D\O\ measurements of $A^s_{SL}$.}
\label{fig:betascomb}
}
\end{figure}

Both the CDF and D\O\  analyses of the \textsf{CP} violation in $B_s^0\to J/\psi\phi$ decay show a slight disagreement with the Standard Model prediction, 
and both results fluctuate in the same direction. Recently the D\O\ collaboration has made public the results of the fit without the strong phase constraints, allowing the HFAG group to combine them with the CDF results~\cite{HFAGsite}. The combined countours are shown in Fig.~\ref{fig:betascomb}. 
The {\it p}-value for the combined result is 3.1\%, corresponding to 2.2 Gaussian standard deviations.

\section{Rare $B_s$ decays at Tevatron and Belle}

\subsection{Leptonic Two-body Decays at Tevatron}

The FCNC process $B^0_{s} \to \mu^+\mu^-$ is predicted to have a branching ratio of 
$\mathcal{B}(B^0_{s} \to \mu^+\mu^-)=(3.42 \pm 0.54)\times 10^{-9}$ in the Standard Model~\cite{Buras:2003td}, well 
below the current experimental sensitivity of the Tevatron experiments.
The $B^0 \to \mu^+\mu^-$ decay is further suppressed by $|V_{td}/V_{ts}|^2$, with a predicted 
branching ratio of $(1.00 \pm 0.14)\times 10^{-10}$.
Significant enhancements are instead predicted by several new physics models.
For example in the minimal super-symmetric standard model (MSSM) the $B^0_s$ branching ratio 
is proportional to $\tan^6\beta$ where $\tan\beta$ is the ratio between the vacuum expectation values
of the two neutral Higgs fields.
In $R$-parity violating super-symmetric (SUSY) models an enhancement is possible even at low 
values of $\tan\beta$.

Both Tevatron experiments have dedicated triggers to collect $B\to\mu^+\mu^-$ events, and 
optimized selections for the $B^0_{s} \to \mu^+\mu^-$ candidates based on sophisticated multivariate 
analysis techniques. D\O\ combines the discriminant variables in a likelihood ratio, while CDF uses a neural network (NN) discriminant.
Both experiments estimate the dominant combinatorial background by a fit to the mass sidebands, while
contribution from decays of B mesons to two light hadrons, which could peak in the signal
mass region, is estimated to be an order of magnitude lower than the combinatorial background.
Both experiments do not report any significant excess of signal candidates over the expected background, 
and set 90\% C.L. limits calculated with a Bayesian method for a data sample of 2 fb$^{-1}$ per
experiment:  $\mathcal{B}(B^0_{s} \to \mu^+\mu^-) < 7.5 \times 10^{-8}$  (D\O\ )~\cite{D0Bsmumu} and 
$\mathcal{B}(B_s^0\to \mu^+\mu^-) < 4.7 \times 10^{-8}$ (CDF)~\cite{Aaltonen:2007kv}.
Because of the superior mass resolution of the tracking system CDF is able to 
separate $B^0_s$ and $B^0$ mesons and to quote
a 90\% C.L. limit separately for the $B^0$ decay of $\mathcal{B}(B^0\to \mu^+\mu^-) < 1.5 \times 10^{-8}$.

Very recently the CDF experiment reported the results of a search for the $B^0_{s,d}\to e^+e^-$ channel and for the 
lepton-flavor violating mode $B^0_{s,d}\to e^+\mu^-$.  In particular the latter decay modes
are strongly suppressed within the Standard Model, in which leptons do not change flavor. These decays are allowed, however, in many models 
of new physics, such as Pati-Salam leptoquarks model, or in SUSY and Extra Dimension models,  where the assumption of
a local gauge symmetry between quarks and leptons at the lepton-flavor violation tree-level couplings leads to the
prediction of a new force of Nature which mediates transitions between quarks and leptons~\cite{pati}.

\begin{figure}[h]
\begin{center}
\includegraphics[width=0.75\textwidth]{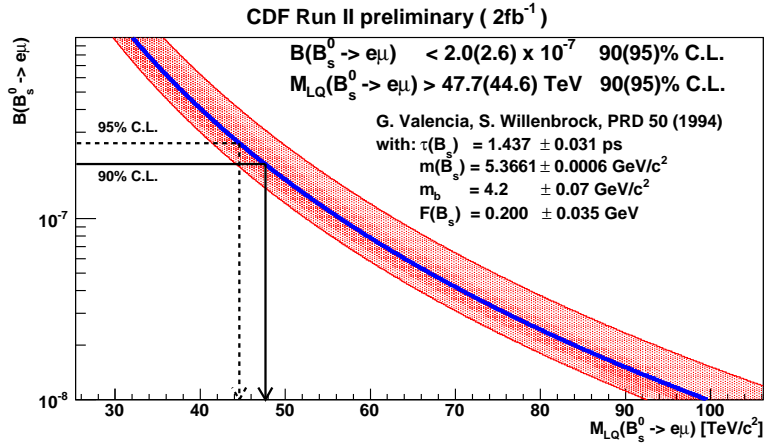}
\end{center}
\caption{Leptoquark mass limit corresponding to the 90\% C.L. on $\mathcal{B}(B_s^0\to e^+\mu^-)$. }
\label{fig:Bsemu}
\end{figure}

Using a 2 fb$^{-1}$ data sample, CDF find one event in the search window for the $B_s^0\to e^+\mu^-$, with estimated 
$0.81\pm0.63$ background events, and one event for  $B_s^0\to e^+e^-$, with $2.8\pm1.8$ estimated background events. 
By using the $B^0\to K^+\pi^-$ decay mode as a relative normalization,  CDF derives the upper limits at 90\% C.L. 
on the decay branching ratios of $\mathcal{B}(B_s^0\to e^+\mu^-) < 2.0 \times 10^{-7}$, and 
$\mathcal{B}(B_s^0\to e^+e^-) < 2.8 \times 10^{-7}$~\cite{Aaltonen:2007kv}.
Finally from the decay branching ratio limits CDF calculate the corresponding lower bound on the Pati-Salam leptoquark mass: 
$M_{LQ}(B_s^0) > 47.7~$TeV/c$^2$ at 90\% C.L. (see Fig.~\ref{fig:Bsemu}).

\subsection{Charmless Two-body Decays at Tevatron}
Non-leptonic decays of $b$ hadrons into pairs of charmless charged hadrons 
are effective probes of the CKM matrix, and sensitive to potential new physics effects.
The large production cross section of $b$ hadrons at Tevatron, and the ability of CDF to trigger 
on fully hadronic decays, allows extending such measurements to \ensuremath{B_{s}^{0}}\ and \ensuremath{\Lambda_{b}^{0}}\ decays,
complementing the \ensuremath{B^{0}}\ meson case, extensively studied at the B-factories.

CDF analyzed an integrated luminosity of $\sim$1 fb$^{-1}$ of pairs of oppositely-charged particles, 
selected by the displaced track trigger.  A sample of 14500 \ensuremath{H^0_{b} \to h^{+}h^{'-}}\ decay modes  
(where  $H^0_{b} = \ensuremath{B^{0}},\ensuremath{B_{s}^{0}} ~{\rm or}~ \ensuremath{\Lambda_{b}^{0}}$  and $h= K,  \pi, p$) 
was reconstructed after the off-line confirmation of trigger requirements.  The invariant mass resolution and the particle identification 
separation power available in CDF, are not sufficient to disentangle the individual $\ensuremath{H^0_{b} \to h^{+}h^{'-}}$ decay modes on an event-by-event basis,
therefore a Maximum Likelihood fit is performed to separate the different components. The fit combines kinematic and particle identification 
information, to statistically determine both the contribution of each mode, and the relative contributions to the \textsf{CP} asymmetries.

Significant signals are seen for \ensuremath{B^0\to \pi^+ \pi^-}, \ensuremath{B^0\to K^+ \pi^-}, and \ensuremath{B_s^0\to K^+ K^-}, previously observed by CDF~\cite{paper_bhh}.
In addition to that, three new rare decay modes have been observed for the first time \ensuremath{B_s^0\to K^- \pi^+}, \ensuremath{\Lambda_b^0\to p\pi^-}\ and \ensuremath{\Lambda_b^0\to pK^-},
with a significance respectively of $8.2 \sigma$, $6.0 \sigma$ and $11.5 \sigma$. No evidence was obtained for \ensuremath{B_s^0\to \pi^+\pi^-} or  \ensuremath{B^0\to K^+K^-}\ mode.

\begin{table}[h]
\centering{
\caption{\label{tab:summary} Branching fractions results. Absolute branching fractions are normalized to the the world--average values
${\mathcal B}(\mbox{\ensuremath{B^0\to K^+ \pi^-}}) = (19.4\pm 0.6) \times 10^{-6}$ and
$f_{s}/f_{d}= 0.276 \pm 0.034$ and $f_{\Lambda}/f_{d}= 0.230 \pm 0.052$~\cite{PDG08}.}
\vspace{0.2cm}
\begin{tabular}{|l|c|}
\hline
Mode          &  \BR (10$^{-6}$)  \\
\hline
\ensuremath{B^0\to \pi^+ \pi^-}        &  5.02 $\pm$ 0.33 $\pm$ 0.35 \\
\ensuremath{B_s^0\to K^+ K^-}          &  24.4 $\pm$ 1.4 $\pm$ 3.5   \\
\hline
\ensuremath{B_s^0\to K^- \pi^+}         &   5.0 $\pm$ 0.7 $\pm$ 0.8   \\
\ensuremath{\Lambda_b^0\to pK^-}          &   5.6 $\pm$ 0.8 $\pm$ 1.5  \\
\ensuremath{\Lambda_b^0\to p\pi^-}         &   3.5 $\pm$ 0.6 $\pm$ 0.9  \\
\hline
\ensuremath{B_s^0\to \pi^+\pi^-}        &  0.49 $\pm$ 0.28 $\pm$ 0.36 \\
 
\ensuremath{B^0\to K^+K^-}          &  0.39 $\pm$ 0.16 $\pm$ 0.12 \\
\hline
\end{tabular}
}
\end{table}

\begin{table}[h]
\centering {
\caption{\label{tab:summary_ACP} \textsf{CP}-violation related results.}
\vspace{0.3cm}
\begin{tabular}{|l|c|}
\hline
Mode & Measurement \\
\hline
$A_{CP}(B^0\to K^+\pi^-)$          & -0.086 $\pm$ 0.023 $\pm$ 0.009         \\
$ A_{CP}(B^0_s\to K^-\pi^+)$     &  0.39  $\pm$ 0.15  $\pm$ 0.08          \\ 
$ A_{CP}(\Lambda_b^-\to pK^-)$        & -0.37  $\pm$ 0.17  $\pm$ 0.03          \\
$ A_{CP}(\Lambda_b^0\to p\pi^-)$      & -0.03  $\pm$ 0.17  $\pm$ 0.05          \\
\hline
\end{tabular}
}
\end{table}

The absolute branching fractions obtained by CDF normalizing the measurements to the the world average value
${\mathcal B}(\mbox{\ensuremath{B^0\to K^+ \pi^-}}) = (19.4\pm 0.6) \times 10^{-6}$, are listed in Table~\ref{tab:summary},  while the CP-related measurements 
are listed in Table~\ref{tab:summary_ACP},  where $f_{d}$, $f_{s}$ and $f_{\Lambda}$ indicate
the production fractions respectively of \ensuremath{B^{0}}\, \ensuremath{B_{s}^{0}}\ and \ensuremath{\Lambda_{b}^{0}}\ from fragmentation of a $b$ quark in $\bar{p}p$ collisions.

The decays \ensuremath{\Lambda_b^0\to p\pi^-} and \ensuremath{\Lambda_b^0\to pK^-} are allowed at tree level in the Standard Model, but are suppressed by the small value
of the involved CKM matrix element $V_{ub}$.
Loop diagram processes can contribute at a magnitude that is comparable to the tree diagram process, leading to 
sizable direct \textsf{CP} violation. In the Standard Model a $A_{CP}$ value of $\mathcal{O}(10\%)$ is predicted.

\begin{figure}[h]
\begin{center}
\includegraphics[width=0.47\textwidth]{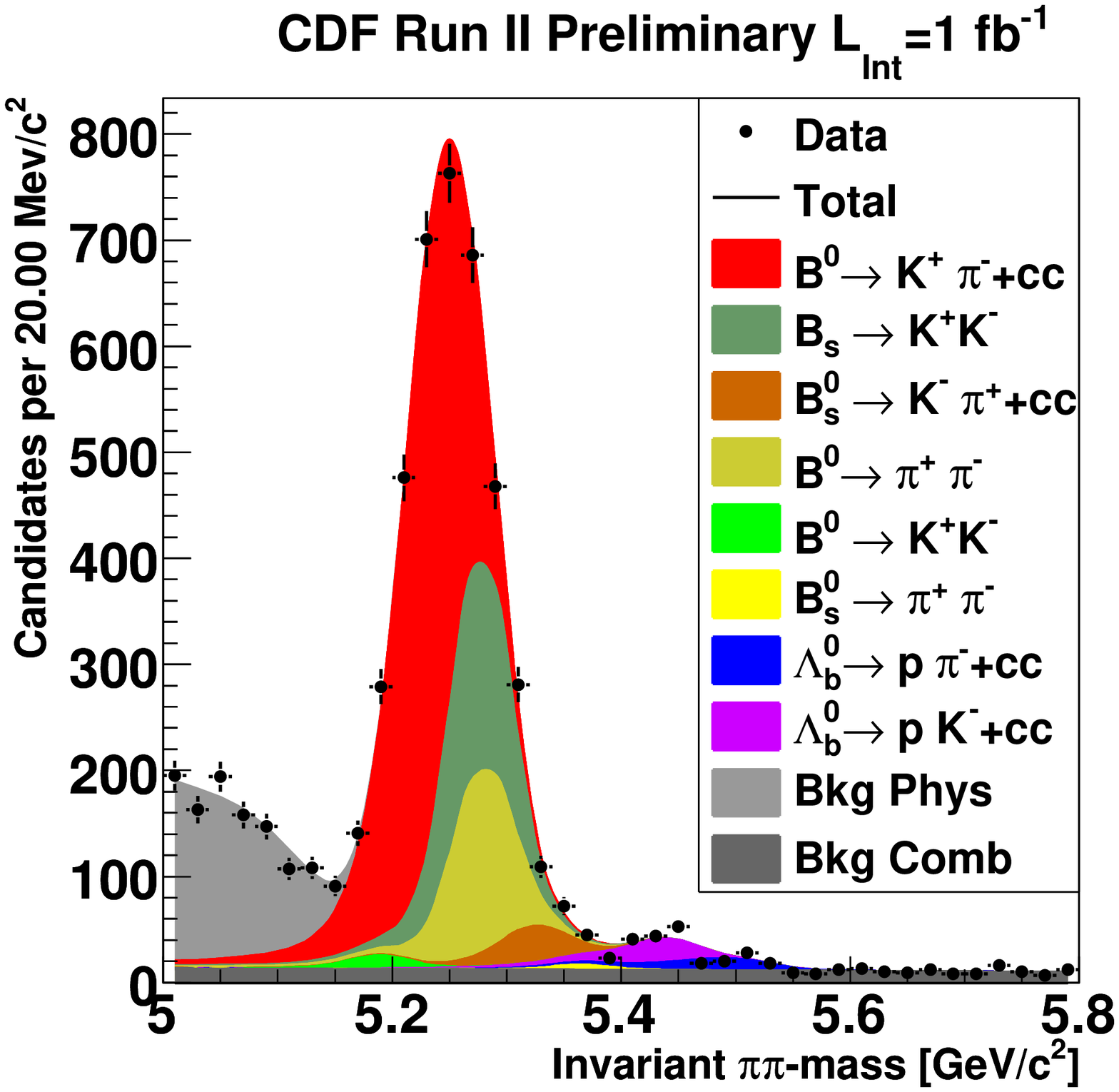}\hspace{0.3cm}
\includegraphics[width=0.47\textwidth]{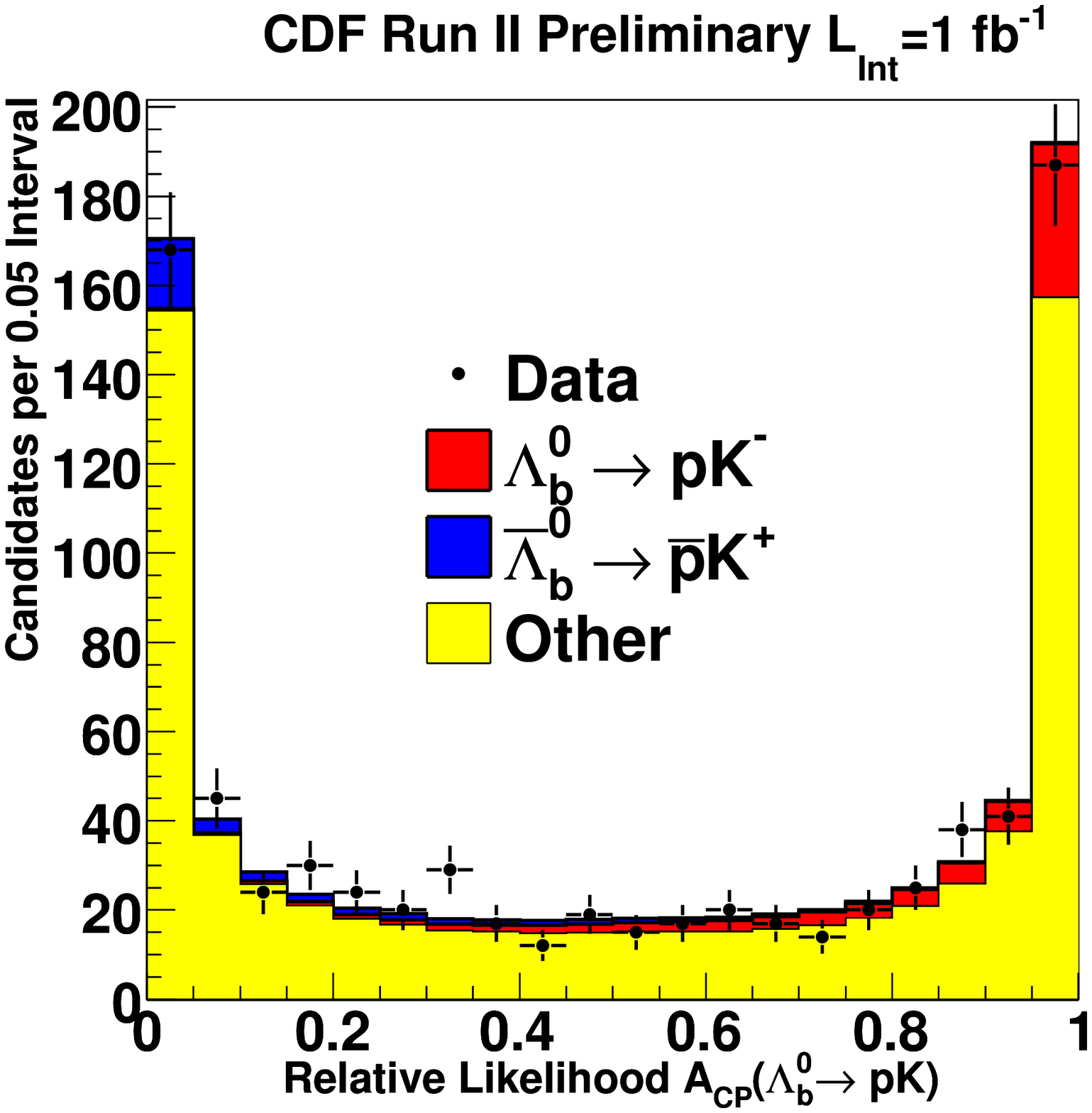}
\end{center}
\caption{Invariant mass spectrum for $\pi^+\pi^-$ mass assignment (left)
and relative probability density function (pdf) of \ensuremath{\Lambda_b^0\to pK^-}:
$\mbox{pdf}(\Lambda_b^0) / [\mbox{pdf}(\Lambda_b^0)+\mbox{pdf}(\bar{\Lambda}_b^0)]$ (right).}
\label{fig:Lbph}
\end{figure}

The measurement of the direct \textsf{CP} violation asymmetries in the $b$-baryon decays, presented by CDF, is the first such measurements in 
this sector. The statistical uncertainty still dominates the resolution and prevents a statement on the presence of asymmetry, whose 
measured value deviates from 0 at 2.1$\sigma$ level in the \ensuremath{\Lambda_b^0\to pK^-}\ decay mode and is fully consistent with 0 in the \ensuremath{\Lambda_b^0\to p\pi^-}\ decay mode. 
In Fig.~\ref{fig:Lbph} the invariant mass spectrum in the $\pi^+\pi^-$ mass hypothesis, and the relative probability density function for the 
$\Lambda_b^0\to pK^-$ are shown, illustrating the good description of the data by the fit and the powerful $\Lambda_b^0$/$\bar{\Lambda}_b^0$ separation.

\subsection{Radiative $B_s$ penguins at Belle}

During the last several years the possibility of performing $B_s^0$ meson studies at the 
$e^+ e^-$ colliders running at the $\Upsilon$(5S) resonance has been extensively explored. The first evidence for $B_s^0$ production at the
$\Upsilon$(5S) was found by the CLEO collaboration \cite{cleoi,cleoe} using
a data sample of 0.42\,fb$^{-1}$ collected in 2003. This study indicated that practical $B_s^0$ measurements at the $\Upsilon$(5S) are possible
with at least 20\,fb$^{-1}$, which can be easily collected at B-factories running with $\sim$10$^{34}$cm$^{-2}$sec$^{-1}$ luminosity. 
To test the feasibility of a $B_s^0$ physics program the Belle collaboration collected at the $\Upsilon$(5S) a sample of 1.86\,fb$^{-1}$ of data in 2005, and 
a one of 21.7\,fb$^{-1}$ in 2006~\cite{beli,bele}.

The collected samples have been used by Belle to perform searches for exclusive radiative decays of the $B_s^0$ mesons, 
that are of great interest, because sensitive to new physics effects and experimentally unaccessible to the Tevatron experiments ,(the presence of low energy photons in the final state makes these kind of decays too hard to be reconstructed in CDF and D\O\ ). 
In particular Belle searched for the decay modes $B_s^0 \to \phi \gamma$ and $B_s^0 \to \gamma \gamma$, using the full 23.6\,fb$^{-1}$ available data sample~\cite{bfiga}.

\begin{figure}[h]
\centering{
\includegraphics[width=30mm]{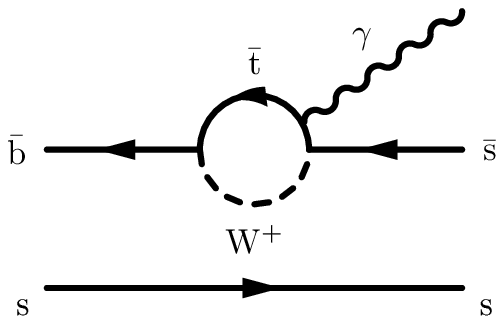}\hspace{2cm}\includegraphics[width=30mm]{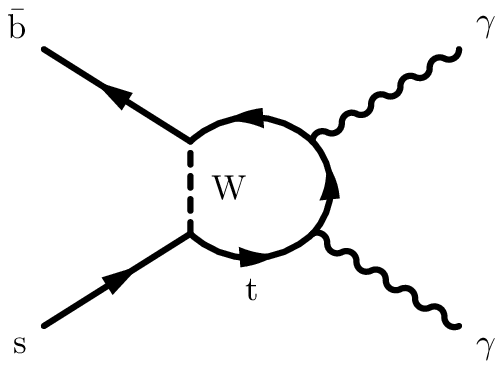}
\vspace{-0.1cm}
\caption{Diagrams describing the dominant SM processes for the
$B_s^0 \to \phi \gamma$ (left) and $B_s^0 \to \gamma \gamma$ (right) decays.}
\label{fig:belle1}
}
\end{figure}

\begin{figure}[h]
\centering {
\includegraphics[width=60mm]{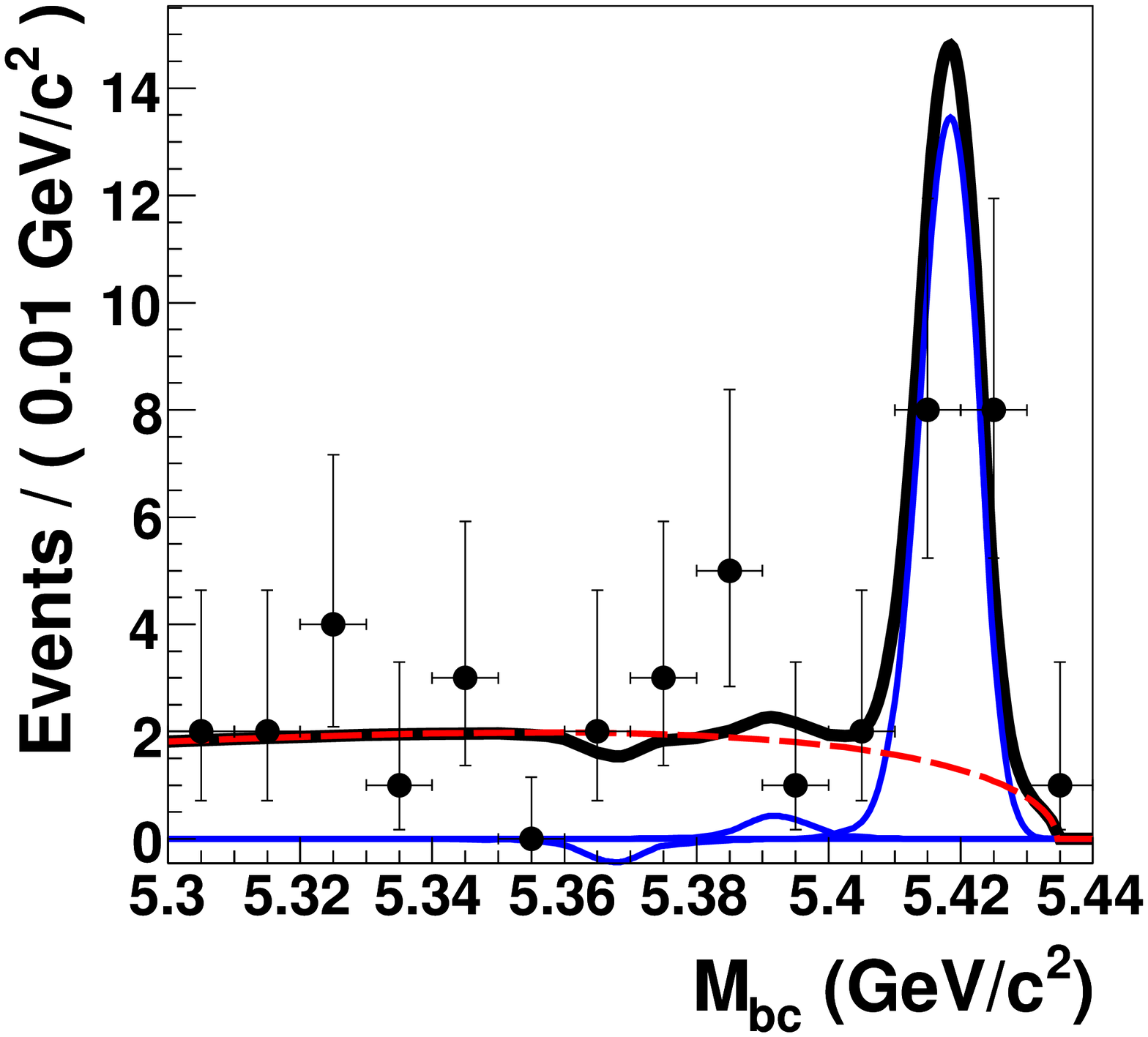}\hspace{0.5cm}\includegraphics[width=60mm]{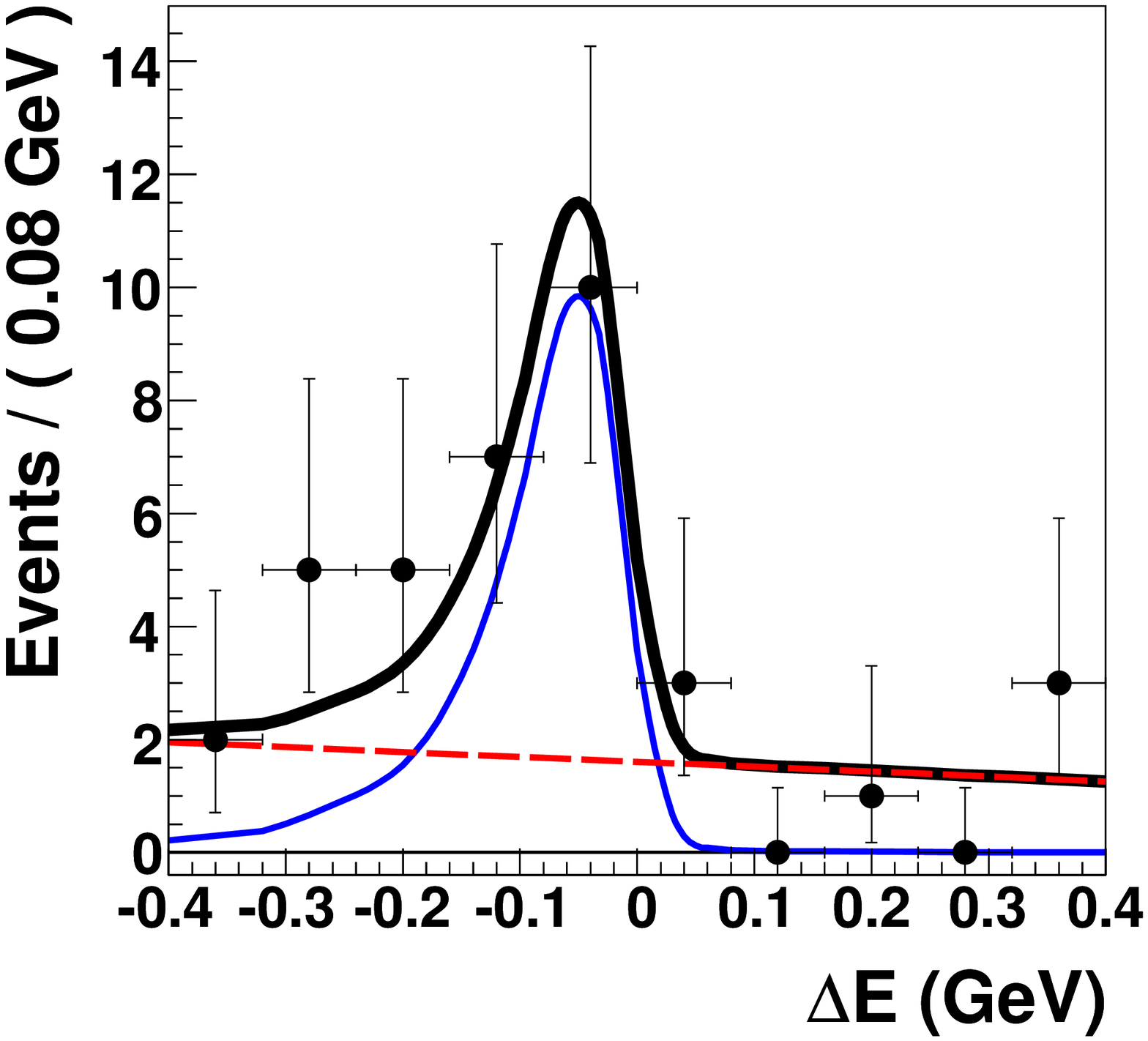} \\
\includegraphics[width=60mm]{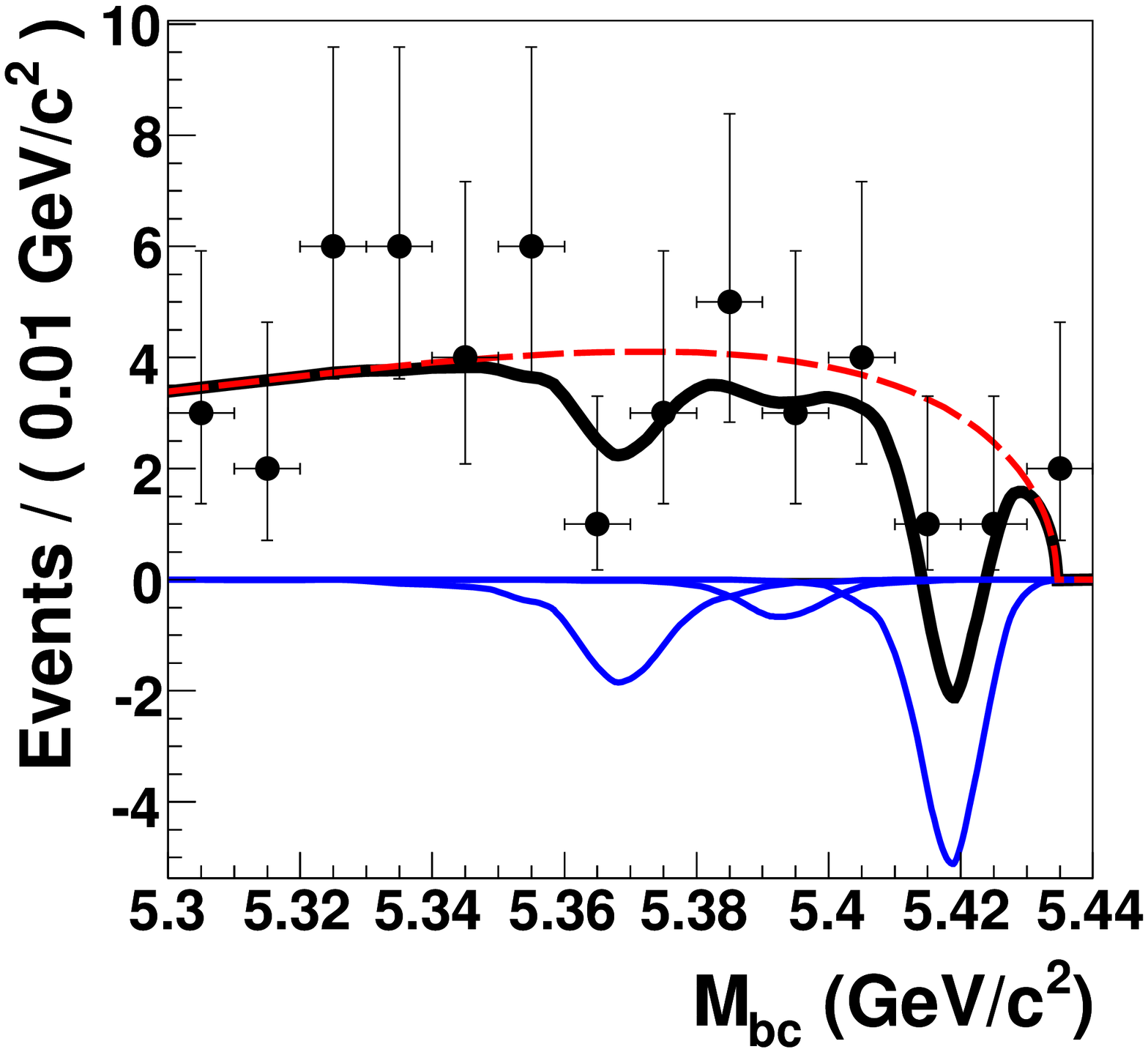}\hspace{0.5cm}\includegraphics[width=60mm]{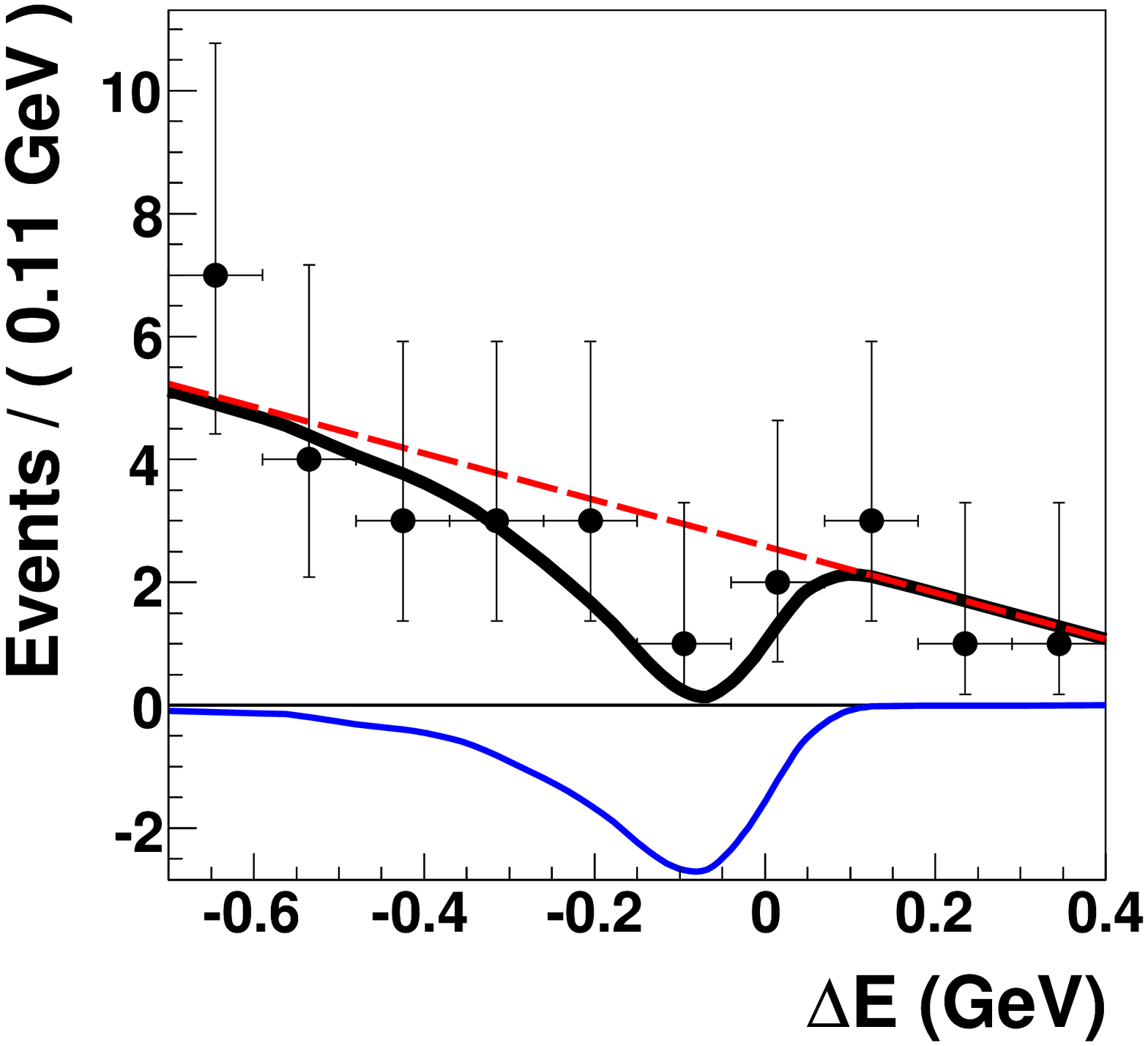}
\vspace{-0.1cm}
\caption{$M_{\rm bc}$ projection (left) and $\Delta E$ projection 
(right) for the $B_s^0 \to \phi \gamma$ (top) and $B_s^0 \to \gamma \gamma$ 
(bottom) modes. The thick solid curves are 
the fit functions (thin solid curves: signal functions,  dashed curves: continuum contribution).}
\label{fig:belle2}
}
\end{figure}
Within the Standard Model (SM) the $B_s^0 \to \phi \gamma$ decay is described by the radiative penguin diagram shown in Fig.~\ref{fig:belle1} (left).
The branching fraction is predicted to be $\sim4 \times 10^{-5}$~\cite{tfigaa}. The $B_s^0 \to \gamma \gamma$ decay proceed via the penguin annihilation diagram shown in Fig.~\ref{fig:belle1} (right) and is expected to have 
a much smaller branching ratio, in the range $(0.5-1.0)\times 10^{-6}$, that can be however enhanced by about an order 
of magnitude in various new physics models~\cite{tfigab,tfigac,tfigad}, reaching a level not far from the current sensitivity of the Belle experiment.

To extract the signal yields a multi-dimensional un-binned extended maximum likelihood fit is performed to the $M_{\rm bc}$ and $\Delta E$ variables.
$M_{\rm bc}$ and $\Delta E$ are respectively the beam-energy-constrained mass, and the energy difference observable, defined 
as: $\Delta E\,=\,E^{CM}_{B_s^0}-E^{\rm CM}_{\rm beam}$ and $M_{\rm bc} = \sqrt{(E^{\rm CM}_{\rm beam})^2\,-\,(p^{\rm CM}_{B_s^0})^2}$,
where $E^{\rm CM}_{B_s^0}$ and $p^{\rm CM}_{B_s^0}$ are the energy and momentum
of the $B_s^0$ candidate in the $e^+ e^-$ center-of-mass (CM) system,
and $E^{\rm CM}_{\rm beam}$ is the CM beam energy.

Fig.~\ref{fig:belle2} shows the $M_{\rm bc}$ and $\Delta E$ projections of the fit results
of the data, together with the fitted functions. A clear signal of $18^{+6}_{-5}$ events is seen in the $B_s^0 \to \phi \gamma$ mode, with a 
significance of 5.5 standard deviations, providing the first observation of a $B_s$ penguin radiative decay.  The branching fraction is measured to be
${\cal B} (B_s^0 \to \phi \gamma) = (5.7^{+1.8}_{-1.5} {\rm (stat.)} ^{+1.2}_{-1.1} {\rm (syst.)}) \times 10^{-5}$,  in agreement
with the SM predictions.

No significant signal is observed instead for the $B_s^0 \to \gamma \gamma$ mode, and an upper limit at the $90\%$ C.L.
of ${\cal B} (B_s^0 \to \phi \gamma) < 8.7 \times 10^{-6}$ is set.

\end{document}